\documentclass[a4paper,11pt]{article}
\usepackage{pos}
\usepackage{subcaption}

\usepackage{color}

\title{Top quark loop corrections to $W^+W^-$ scattering in EChL}

\author{Antonio Dobado}
\author*{Carlos Quezada-Calonge}
\author{Juan José Sanz-Cillero}

\affiliation{Facultad de Ciencias Físicas and IPARCOS, Universidad Complutense de Madrid,\\
  Plaza de las Ciencias 1, Madrid, Spain}

\emailAdd{dobado@fis.ucm.es}
\emailAdd{cquezada@ucm.es}
\emailAdd{jjsanzcillero@ucm.es}

\abstract{
    We calculate fermion-loop corrections to high energy $W^+ W^-$ scattering in the context of a Strongly Interacting Electroweak Symmetry Breaking Sector (EWSBS) using Higgs Effective Field Theory (HEFT). We test the assumption that these corrections are negligible when compared to the boson-loop ones, as it is commonly taken for granted in the literature. While this is correct in most cases, 
    we find that, for some particular regions of the parameter space, fermion-loops can be important: deviations in the couplings of the HEFT from their Standard Model values may lead to fermion-loop corrections as relevant as the boson-loop ones. }

\FullConference{%
  40th International Conference on High Energy physics - ICHEP2020\\
  July 28 - August 6, 2020\\
  Prague, Czech Republic (virtual meeting)}

\begin{document}

\maketitle

\section{Introduction}

A very interesting scenario to  seek for new physics (NP) is the electroweak boson scattering at the LHC. Deviations from the Standard Model (SM) could come from a strongly interacting EWSBS \cite{Delgado:2013loa}, giving rise to a enhancement of the longitudinal components of $W$ and $Z$ interactions at high energies. In this situation the most efficient phenomenological description of the NP is the so called Higgs Effective Field Theory (HEFT), which is a sort of Higgs-equipped (Electroweak) Chiral Lagrangian (EChL) \cite{Appelquist}. At high energies an important element 
for analysis is the Equivalence Theorem (ET)~\cite{ET} which relates, up to corrections $\mathcal{O}(M_W/ \sqrt{s})$, longitudinal electroweak (EW) gauge bosons and  would-be Goldstone-bosons (WBGB) amplitudes thus largely simplifying computations.
In this context usually only  boson-loop corrections are taken into account but, 
 as fermion-loop corrections formally scale like $\mathcal{O}(M_{\rm fer}^2 s)$, the latter are regularly ignored.
 
 In this note we report on  some of our first results concerning a systematical quantitative study of these fermion-loop contributions to the $W$ and $Z$ boson scattering at the high energies relevant for the LHC. It is well known that  fermion-loops are proportional to the mass of the fermion inside the loop and to its couplings to other particles (some of them may still have a  10 \% of deviation with respect to the SM values \cite{higgsxsection}). To start with, we will focus on the clearly dominant top-quark  corrections and we will test their relevance considering the whole range of phenomenologically possible coupling values. We will see that  for some values of the HEFT parameters space, these contributions can be significant. The precise details of our computations cannot be described in this short note and will be given elsewhere~\cite{in-preparation}.

 

%

 In order to check the relative importance of fermion-loop compared with boson-loops we will focus on the imaginary part of the amplitudes since they enter at next-to-leading order in the chiral counting and thus are not masked by the purely real lowest-order amplitude. In particular we  will concentrate in the ratio fermion to boson plus fermion contribution for the first two  partial wave amplitudes (PWA)  ($J=0$ and $J=1$). In these preliminary results only longitudinal electroweak bosons have been considered.. 

\section{The top quark in the EChL }

At leading order (LO), the relevant part of our effective Lagrangian is given by:
\begin{equation}
\mathcal{L}_S  = \frac{1}{2} \mathcal{F}(h) \partial_\mu \omega ^i \partial^\mu \omega_j \left( \delta_{ij}+\frac{\omega_i \omega_ j}{v^2}\right)  +\frac{1}{2}  \partial_\mu h \partial^\mu h  - V(h)+ \mathcal{L}_Y,
\end{equation}
with the $\mathcal{L}_Y$ providing the Yukawa interactions between fermions and scalars ($h$ is the Higgs and $\omega^a$ the WBGB 
fields). 
In the limit $M_b \ll M_t$ the interactions between scalars and the bottom quark can be neglected, and the LO coupling of the top  to the HEFT is given by the Lagrangian:
\begin{equation}
 \mathcal{L}_Y  =-\mathcal{G}(h) \bigg [\sqrt{ 1-\frac{\omega^2}{v^2}}M_t t \bar{t}+i\frac{\omega ^0}{v} M_t \bar{t}\gamma^5t  -i\frac{ \sqrt{2} \omega ^+}{v}M_t \bar{t} P_L b +
   i\frac{ \sqrt{2} \omega ^-}{v}M_t \bar{b} P_R t \bigg ].
\end{equation}
where $P_{R,L}=\frac{1}{2}(1\pm \gamma_5)$ and we introduce the Higgs functions:
\begin{equation}
    \mathcal{G}(h)=1+{c_1} \frac{h}{v}+{c_2} \frac{h^2}{v^2}+..., \quad  \mathcal{F}(h)=1+2{a} \frac{h}{v}+{b} \frac{h^2}{v^2}+... \quad\text{and}\quad V(h)= \frac{M_h^2}{2}  h^2+ d_3 \frac{M_h^2}{2v^2}  h^3+...
\end{equation}


%

In the SM case one has $a=b=c_1=d_3=1$ and $c_2=0$. Ultimately, for a study beyond the ET, one must also add the EW gauge boson interactions to the EChL~\cite{Appelquist}.

\vspace*{-0.3cm}
\section{Loop corrections to elastic $W^+W^-$ scattering}

At LO, $ O(p^2)$, this amplitude $T_2$ is purely real. Next contribution $T_4$ shows up at $ O(p^4)$ and consists of a real tree-level part  $T_{4,\rm tree}$ and one-loop diagrams 
giving the $T_{4,1\ell}$ amplitude an imaginary part. Up to the order studied in this work, $O(p^4)$, the real part of the amplitude is provided by the mentioned three contributions, $\mbox{Re}T=T_2+T_{4\rm tree}+ \mbox{Re}T_{4,1\ell}$.  
This makes the study of the full NLO one-loop corrections cumbersome. On the other hand, the imaginary part only gets contributions from one-loop diagrams up to this order, $\mbox{Im}T=\mbox{Im}T_{4,1\ell}$. 


We will study the imaginary part of the PWA $a_J(s)$, provided by the decomposition  
$T(s,t)=\sum_J 16\pi K (2J+1) P_J(x) \, a_J(s)$, with $x=\cos\theta$ and $K=1$ ($K=2$)  for distinguishable (indistinguishable) final particles. In the physical energy region, Im $a_J(s)$ will be obtained from the one-loop absorptive cuts in the $s$-channel, which we will use as a measure of the relative importance of the various contributions.

For scattering amplitudes with only bosons as external legs it is possible to clearly separate fermion and boson loop diagrams (no mixed loops appear). We use the following notation to refer to the corresponding absorptive cuts: ${\rm Fer}_J=Im[a_J]_{t\bar{t}}$ and ${\rm Bos}_J=Im[a_J]_{WW,ZZ,HH}$ .



The absorptive cuts for longitudinal $W^+W^-$ elastic scattering can be found in Ref.~\cite{espriu} and  for $ZZ$ and $HH$ in Ref.~\cite{Delgado:2013loa}. For the case of absorptive cuts with intermediate EW bosons, only longitudinal polarizations are being considered.
This is because  we are interested in scenarios with an EWSBS where longitudinal components dominate the high energy dynamics.

It is important to notice how the relevant couplings enter in each PWA. For $J=0$, \ $\rm{Fer_0}$ depends on $a,c_1$ and $ \rm Bos_0$ on $a,b,d_3$ while, for $J=1$, $\rm Bos_1$ depends only on $a$ and  $\rm Fer_1 $ has no dependence (only $M_t$ and $v$).


The goal of this note is to point out that there are regions of the parameter phase-space where fermion loops become as important as 
the bosonic loops and thus they should not be neglected. Since by unitary the imaginary parts of the PWA are always positive we will consider the ratio :
\begin{equation} R_J=\frac{{\rm Fer}_J}{{\rm Bos}_J+{\rm Fer}_J} \,\, \in\,\, [0,1]\,. \end{equation}
Values close to 0 will indicate we can safely drop the top quark contribution while significant deviations from 0 will point out their relevance.




\vspace*{-0.3cm}
\section{Results for $W^+W^-$ scattering}

We have explored the couplings in the phenomenological admissible range  \cite{higgsxsection}, this is: $0.9\leq \, a,b,c_1 , d_3\,\leq 1.1$. In the following plots we have scanned one coupling at a time while keeping the others fixed to their SM values for reference. Concerning the center-of-mass energy we have considered the interval  $0.5$~TeV$\leq \sqrt{s}\leq 3$~TeV, which is the relevant one to look for NP at the LHC.


\vspace*{-0.2cm}
\subsection{$R_0$}

At this point it is important to state that, when dealing with values of the parameters close to the SM, the ET  is no longer so useful at the energies considered. This is because the SM is a renormalizable theory where the longitudinal components of the EW bosons are not strongly interacting and do not play a dominant role~\cite{Delgado:2013loa}. 
Indeed, the scattering amplitude between two scalars vanishes in the zero-mass limit $M_h,\, M_W,\, M_Z\to 0$. 
Thus one has to go beyond the ET in this case. This has been done for the $W^+W^-$ cuts while, in these preliminary results, we have only used the ET for the $ZZ$ and $HH$ cuts.

\begin{figure}[!t]
\centering
\begin{subfigure}{0.4 \columnwidth}
  \centering
  \includegraphics[width=1\linewidth]{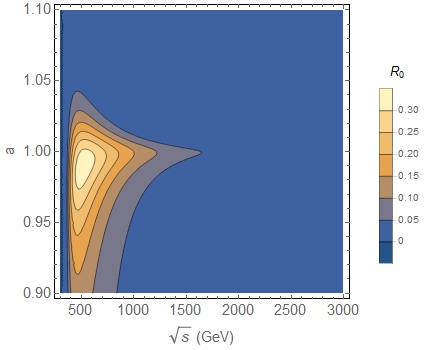}  
  \caption{ $R_0$ dependance on $a$ for $b=c_1=d_3=1$.}
  \label{fig:r0chnga}
\end{subfigure}
\hspace*{0.75cm}
\begin{subfigure}{0.4 \columnwidth}
  \centering
  \includegraphics[width=1 \linewidth]{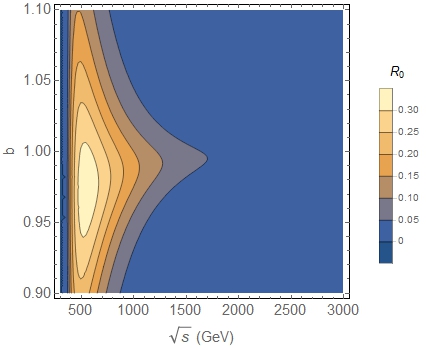}  
  \caption{$R_0$ dependance on $b$ for $a=c_1=d_3=1$.}
  \label{fig:r0chngb}
\end{subfigure}
%
\\[8pt]
\centering
\begin{subfigure}{0.4 \columnwidth}
  \centering
  \includegraphics[width=\linewidth]{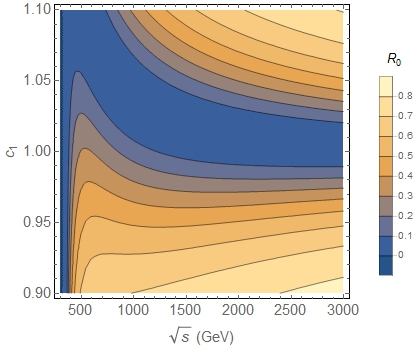}  
  \caption{$R_0$ dependance on $c_1$ for $a=b=d_3=1$.}
  \label{fig:r0chngc1}
\end{subfigure}
\hspace*{0.75cm}
\begin{subfigure}{0.4 \columnwidth}
  \centering
  \includegraphics[width=\linewidth]{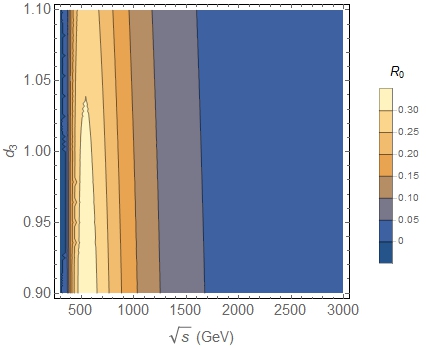}  
  \caption{$R_0$ dependance $d_3$  for $a=b=c_1=1$.}
  \label{fig:r0chngd3}
\end{subfigure}
\label{fig:fig1}
\caption{}
\end{figure}

As one can see in Figure \ref{fig:r0chnga} and Figure \ref{fig:r0chngb}, when we scan $a$ and $b$ we can find order 30 $\%$ corrections around 0.5 TeV. As the energy increases fermion corrections rapidly become irrelevant, as expected.

In the  $c_1$ case (Figure \ref{fig:r0chngc1}), $R_0$ ranges from 10 to 80 \% at 0.5 TeV and, as the energy increases, values of $c_1$ around the SM yield a smaller $R_0$. The dependence on $d_3$ is negligible as we see in Figure $\ref{fig:r0chngd3}$ but we find corrections of order 30 \% at low energies. 
If we do a parameter scan for all values between 0.9 and 1.1 for two benchmark energies, 1.5 TeV and 3 TeV, the maximum $R_0$ happens for $c_1=0.9$ and all other parameters set to the SM.

\vspace*{-0.2cm}
\subsection{$R_1$}

\begin{figure}[h]
\includegraphics[width=.4 \linewidth]{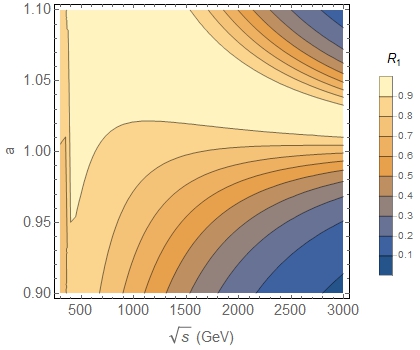}
\centering
\caption{$R_1$ dependance on a}
\label{fig:r1chnga}
\end{figure}

For the $J=1$ PWA,  Fer$_1$ does not depend on any parameter apart of the mass of the top quark, while the boson part Bos$_1$ depends only on $a$ through the $W^+W^-$ cut. From  Figure \ref{fig:r1chnga} we find a wide range of corrections for low energies ( 80-90 \% at 0.5 TeV for the full range of $a$) and for high energies (10-90 \% at 3 TeV in the whole range). Thus, in this case the assumption that fermion corrections can be neglected does not hold for our study of the imaginary part of the partial waves.

\subsection{Specific Scenarios: Minimal Composite Higgs Model (MCHM)}

From the previous plots we can see that, when the values of some parameters are different from 1, fermion corrections can be in fact relevant. This is the case for example of some NP scenarios like the MCHM where the parameters depend on the NP scale $f$. Choosing a value for $a=a^*$ we can find the scale of NP $f$ via $a^*=c_1^*=d_3^*=\sqrt{1-\xi}$ and  $b^*=1-2 \xi$, with $\xi=v^2/f^2$~\cite{MCHM}.

\begin{figure}[!h]
\begin{subfigure}{0.5 \columnwidth}
  \centering
  \includegraphics[width=\linewidth]{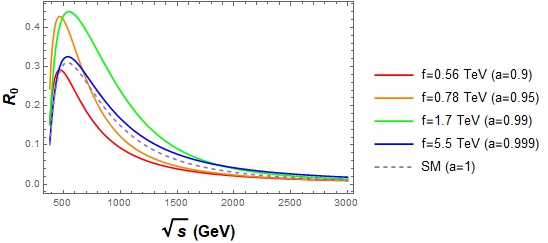}  
  \caption{$R_0$ ratio for the $J=0$ PWA in the MCHM}
  \label{fig:r0composite}
\end{subfigure}
\begin{subfigure}{0.5 \columnwidth}
  \centering
  \includegraphics[width=\linewidth]{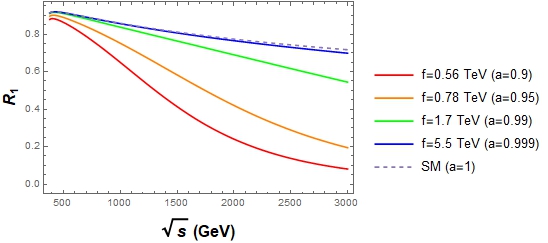}  
  \caption{$R_1$ ratio for the $J=1$ PWA in the MCHM}
  \label{fig:r1composite}
\end{subfigure}
\caption{}
\label{fig:MCHM}
\end{figure}

As seen in Figure \ref{fig:r0composite}, $R_0$ has a maximum of 45 \% for $a=0.9$   around 0.6~TeV and rapidly decreases. For values closer to the SM corrections are even smaller. For $R_1$ the opposite happens;  close to the SM values, corrections are important at low and high energy. For example, for $a=0.99$ fermion corrections move between 90 \% and 60 \%. Again, $J=1$ is more sensitive to fermion loops.

\vspace*{-0.25cm}
\section{Conclusions}

We have studied the top quark loop contribution to $W^+ W^- \rightarrow W^+ W^-$ amplitude using the EChL and compared it with the boson ones. Though boson contributions dominate in most of the parameter space,   there are small regions where fermions become relevant. For the $J=1$ partial-wave ratio $R_1$, fermion corrections could be as important, or even greater, than the boson ones. As we showed, the most important parameter for the $J=0$ ratio $R_0$ is $c_1$. In this preliminary analysis we find the largest correction for $a=b=d_3=1$ and $c_1=0.9$ for energies between 1.5 and 3 TeV, obtaining a 84\% contribution for the latter. 
%
%
For $R_1$ (which only depends on $a$), we find that, even for values close to the SM, fermion corrections are in fact relevant, going all the way up to 90 \%. Regarding  the MCHM, the $R_1$ ratio moves from 90 \% (close to the SM) to 20 \% ($a=0.9)$. $R_0$ has a maximum of 40 \% correction at low energies (0.5 GeV) when $a=0.9$. Therefore $R_1$ is the most sensitive channel to test fermion corrections.

In future work all the intermediate states and polarizations (not only longitudinal) will be included, and we will study their relevance~\cite{in-preparation}.  Finally, given the possibility of a strongly interactiong EWSBS, one should deal with the unitarity problem of the perturbative amplitude~\cite{Dobado:2019fxe}.


\vspace*{-0.25cm}
\section*{Acknowledgements} We would like to thank our collaborators A. Castillo, R. L. Delgado and F. Llanes-Estrada,  who participated in the earlier parts of the research presented in this note \cite{Castillo:2016erh}. 
This research is partly supported by the Ministerio de Ciencia e Inovaci\'on under research grants FPA2016-75654-C2-1-P and PID2019-108655GB-I00; by the EU STRONG-2020 project under the program H2020-INFRAIA-2018-1 [grant agreement no. 824093]; and by the STSM Grant from COST Action CA16108. C. Quezada-Calonge has been funded by the MINECO (Spain) predoctoral grant BES-2017-082408.

\end{document}